\documentclass{PoS}

\title{Measuring the Weizs\"acker-Williams distribution of linearly polarized gluons at an electron-ion collider through dijet azimuthal asymmetries}

\ShortTitle{Weizs\"acker-Williams distribution of linearly polarized gluons at an EIC }

\author{\speaker{Vladimir Skokov}\thanks{
This research was supported in part by the ExtreMe Matter Institute EMMI at the GSI Helmholtzzentrum fuer Schwerionenphysik, Darmstadt, Germany.
}\\
Department of Physics, North Carolina State University, Raleigh, North Carolina 27695, USA\\
RIKEN-BNL Research Center, Brookhaven National Laboratory, Upton, New York 11973-5000, USA
        \\
        E-mail: \email{VSkokov@ncsu.edu}}

\abstract{
The  Weizs\"acker-Williams  transverse momentum dependent (TMD)   gluon  distributions can be probed
in the production of a hard dijet  in semi-inclusive DIS.  
This process is sensitive  not only to the conventional but also to the 
linearly polarized gluon distribution. 
The latter gives rise to an azimuthal dependence of the dijet cross section and therefore can be distinguished from the former. 
Feasibility study of a measurement of these TMDs through dijet production at a future electron-ion collider
shows that the extraction of the distribution of linearly polarized gluons with a 
statistical accuracy of 5\%   will require  estimated  luminosity of 20 fb$^{-1}$/A.
}

\FullConference{XXVII International Workshop on Deep-Inelastic Scattering and Related Subjects - DIS2019\\
		8-12 April, 2019\\
		Torino, Italy}

\begin{document}

\section{Introduction}

In this proceedings, I summarize the fundings of Ref.~\cite{Dumitru:2018kuw}, where 
the conventional and linearly polarized
Weizs\"acker-Williams (WW) gluon distributions at small
$x$~\cite{Dominguez:2011wm,Dominguez:2011br} were studied with the goal of accessing the feasibility of a measurement of
the gluon distributions at an EIC through the dijet production
process.

At leading order in $\alpha_s$ and 
in the small-$x$, high-energy limit, to leading power in the inverse dijet total transverse momentum,
the cross-section for inclusivexproduction of a $q+\bar q$ dijet in high energy deep inelastic
scattering of a virtual photon $\gamma^*$ off a proton or nucleus is
given by~\cite{Dominguez:2011wm,Metz:2011wb}
\begin{eqnarray}
E_1E_2
\frac{d\sigma ^{\gamma _{T}^{\ast }A\rightarrow
    q\bar{q}X}}{d^3k_1d^3k_2 d^2b}
&=&\alpha _{em}e_{q}^{2}\alpha _{s}\delta \left( 1 -z-\bar z\right) z\bar z\left( z^{2}+\bar z^{2}\right) \frac{\epsilon _{f}^{4}+
{P}_{\perp }^{4}}{({P}_{\perp }^{2}+\epsilon _{f}^{2})^{4}}
\nonumber \\
&&
\quad \quad \quad \quad \quad \quad
\times \left[ xG^{(1)}(x,q_{\perp })-\frac{2\epsilon _{f}^{2}{P}%
_{\perp }^{2}}{\epsilon _{f}^{4}+{P}_{\perp }^{4}}\cos
  \left(2\phi\right)xh_{\perp }^{(1)}(x,q_{\perp })\right] ~,
\label{eq:dijet_T} \\
E_1E_2
\frac{d\sigma ^{\gamma _{L}^{\ast }A\rightarrow q\bar{q}X}}{d^3k_1d^3k_2 d^2b}
&=&\alpha _{em}e_{q}^{2}\alpha _{s}\delta \left( 1 -z-\bar z\right) z^2\bar z^2\frac{8\epsilon _{f}^{2}{P}_{\perp }^{2}}{(
{P}_{\perp }^{2}+\epsilon _{f}^{2})^{4}}  \nonumber \\
&&
\quad \quad \quad \quad \quad \quad
\times \left[ xG^{(1)}(x,q_{\perp })+\cos \left(2
  \phi\right)xh_{\perp }^{(1)}(x,q_{\perp })\right]~, 
\label{eq:dijet_L}
\end{eqnarray}
where  $b$ is the impact parameter. 
The transverse momenta (light-cone momentum fractions) of the produced quark and anti-quark are given
by $\vec{k}_{1\perp}$ ($z$)  and $\vec{k}_{2\perp}$ ($\bar z$). 
These quantities can be combined into the dijet total  transverse momentum $\vec{P}_{\perp }$ and
the momentum imbalance $\vec{q}_\perp$: 
\begin{equation}
\vec{P}_{\perp } = \bar z \vec{k}_{1\perp} - z \vec{k}_{2\perp}~~,~~
\vec q_\perp = \vec{k}_{1\perp}+\vec{k}_{2\perp}\, .
\label{eq:Ptqtdef}
\end{equation}
The angle $\phi$ denotes the azimuthal
angle between $\vec{P}_{\perp }$ and $\vec{q}_{\perp }$.
 Only the case when $\vec{P}_{\perp }$ is greater than
$\vec{q}_\perp$, also known as the ``correlation
limit'', is considered here.  
Power corrections to Eqs.~(\ref{eq:dijet_T},\ref{eq:dijet_L}) generate additional contributions 
$\sim (Q_s^2/P_\perp^2) \log P_\perp$ to the isotropic and $\sim \cos
2\phi$ terms~\cite{Dumitru:2016jku}.
Also,  a $\cos 4\phi$ angular
dependence arises from power corrections of order
$q_\perp^2/P_\perp^2$.
Although these corrections might be important for phenomenology at an EIC, 
they are neglected in the present summary.  

The average $\cos 2\phi$ measures the azimuthal anisotropy,
$v_2 \equiv \langle \cos 2 \phi\rangle$, where averaging is performed  over $\phi$ at 
  fixed $q_\perp$ and $P_\perp$, with
normalized weights proportional to the respective cross-sections.

The gluon $x$, 
\begin{equation} \label{eq:x_g}
x = \frac{1}{W^2+Q^2-M^2}\left( Q^2 +
q_\perp^2 + \frac{1}{z\bar z}P_\perp^2\right)\,,
\end{equation}
is independent of $\phi$ and, therefore, for definite polarization of the virtual
photon we obtain~\cite{Dumitru:2015gaa}
\begin{equation} \label{eq:v2_L_T}
v_2^L = \frac{1}{2} \frac{xh_{\perp }^{(1)}(x,q_{\perp
  })}{xG^{(1)}(x,q_{\perp })}~~~,~~~
v_2^T = - \frac{\epsilon _{f}^{2}{P}_{\perp }^{2}}{\epsilon
  _{f}^{4}+{P}_{\perp }^{4}} \frac{xh_{\perp }^{(1)}(x,q_{\perp
  })}{xG^{(1)}(x,q_{\perp })}~.
\end{equation} 
In experiments it is
not possible to distinguish the polarization of the photon in dijet
production. Nevertheless, 
 as will be shown at the end
of this proceedings, 
a careful analysis of experimental data
may give access to contributions from different polarizations. 

An important phenomenological difference between the conventional and 
linearly polarized distribution is that the former   can also be measured in $\gamma A \to q\bar q
X$ in the $Q^2\to0$ limit, while  for a real photon
$\epsilon_f^2 \propto Q^2 \to0$ the cross-section for the
dijet production becomes isotropic and no longer useful for extracting $xh_{\perp
}^{(1)}(x,q^2_\perp)$.

\begin{figure}[t]
	\centerline{
\includegraphics[width=0.5\linewidth]{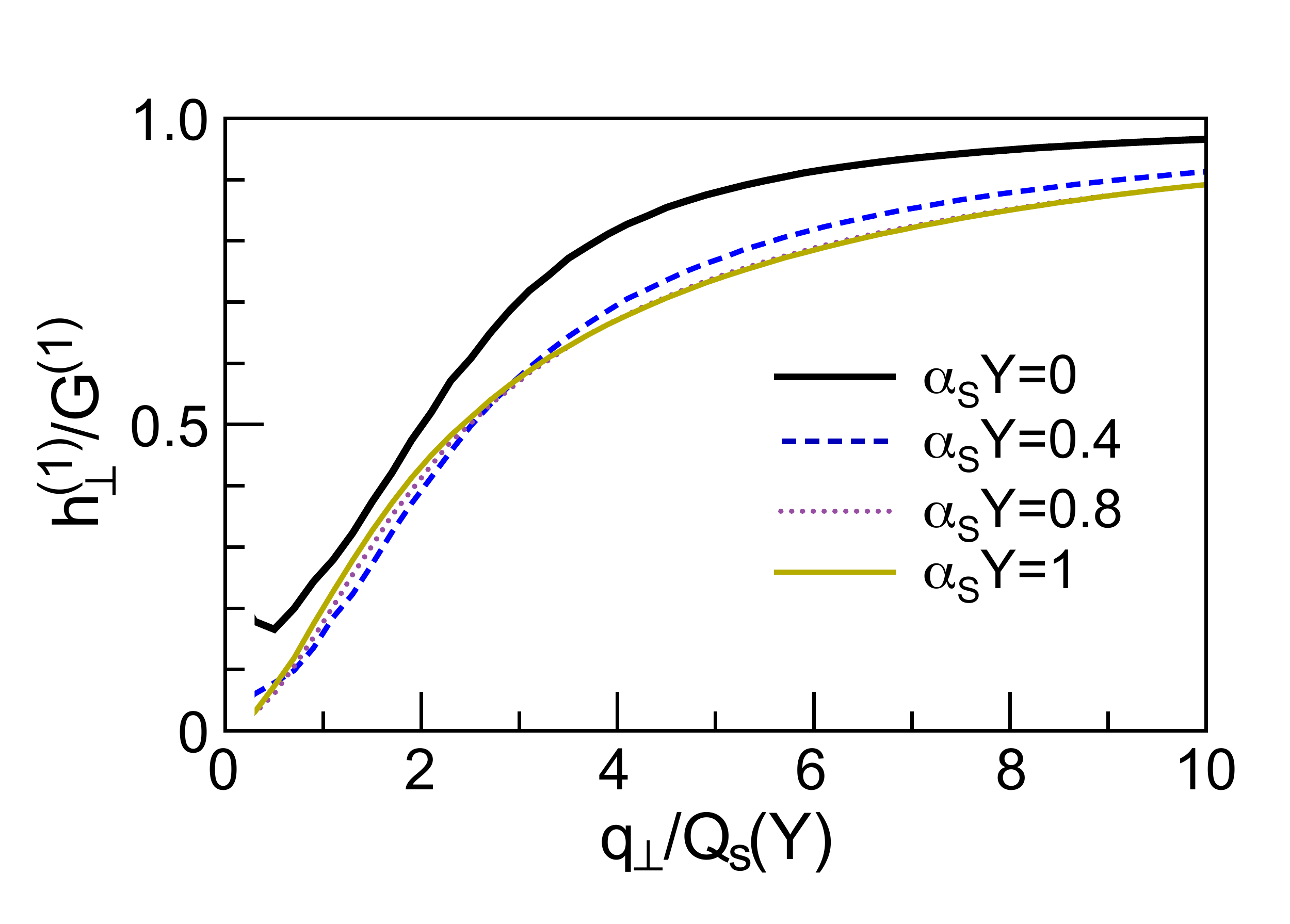}
}
\caption{$xG^{(1)}(x,q^2_\perp)$ and $xh^{(1)}(x,q^2_\perp)$ versus transverse momentum $q_\perp$ at different rapidities
  $Y=\log x_0/x$. $Q_s(Y)$ is the saturation momentum. 
  }
\label{fig:xGxh}       
\end{figure}
Numerical solutions of
the JIMWLK evolution equation to small $x$ were presented in Ref.~\cite{Dumitru:2015gaa}, shown
in Fig.~\ref{fig:xGxh}.
At small transverse momentum, the polarization is significantly suppressed. At  $q_\perp\gg Q_s(Y) $, 
$xh^{(1)}(x,q^2_\perp)\to xG^{(1)}(x,q^2_\perp)$ corresponding to
maximal polarization.
 For the momentum imbalance of order the saturation momentum,    these numerical
simulations show a substantial angular modulation of the dijet cross-section, because    $xh^{(1)}(x,q^2_\perp)/ xG^{(1)}(x,q^2_\perp)\simeq 10\%
- 20\%$.

To simulate $q+\bar q$ dijet production, described by  Eqs.~(\ref{eq:dijet_T}) or~(\ref{eq:dijet_L}), a Monte-Carlo code (MCDijet) was developed in Ref.~\cite{Dumitru:2018kuw}. 
For details on the implementation,  we refer the reader to  Ref.~\cite{Dumitru:2018kuw}, instead we turn to practical application of  MCDijet to an EIC.  

\section{Feasibility studies} 

In order to 
show that the anisotropy generated on  $q+\bar q$ level is not lost during  
reconstruction of  dijets within restrictions of a realistic
detector environment  and to estimate that the DIS background processes can
be suppressed sufficiently by kinematic  cuts  not to affect the level of anisotropy, we performed the analysis of pseudo-data generated by the
Monte Carlo generator MCDijet, PYTHIA 8.2 \cite{Sjostrand:2014zea} for
showering of partons generated by MCDijet, and PYTHIA 6.4
\cite{Sjostrand:2006za} for background studies. Jets are reconstructed
with the  FastJet package~\cite{Cacciari:2011ma}.

\begin{figure}[h]
\centering \includegraphics[width=0.85\linewidth]{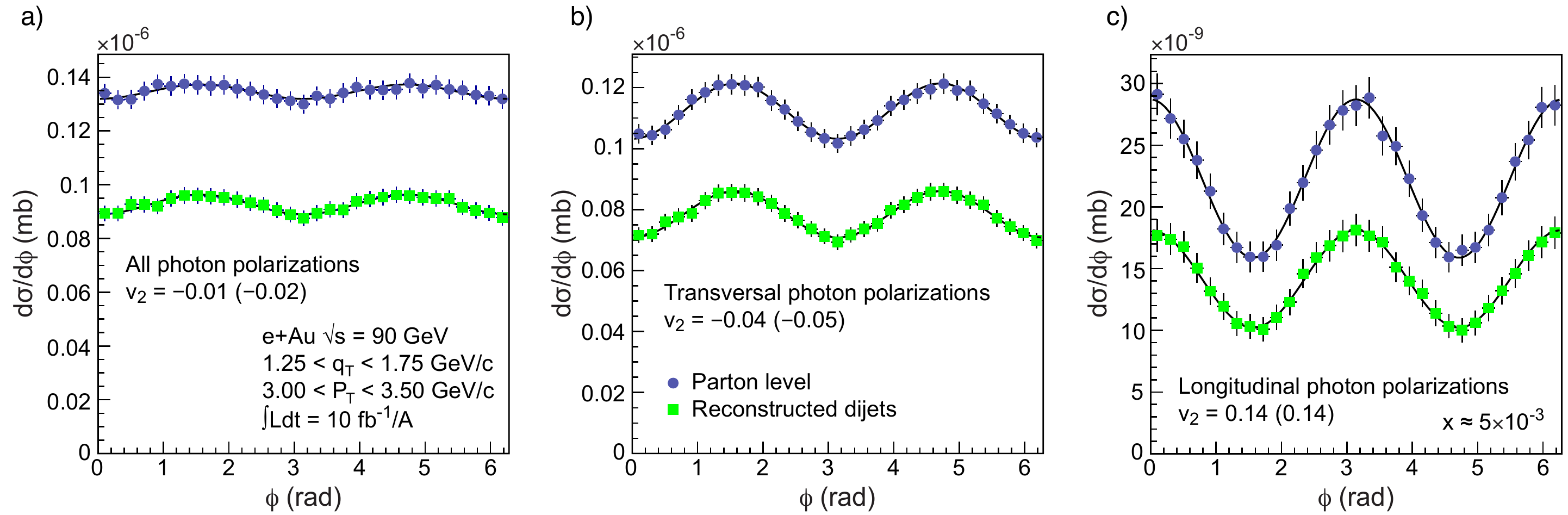}
\caption{$\protect\mathrm{d}\sigma/\protect\mathrm{d}\phi$
  distributions for parton pairs (circles) generated with the
  MCDijet generator and corresponding reconstructed dijets (squares) in $\sqrt{s}$=90 GeV $e$+A collisions for $1.25 < q_\perp <
  1.75$ GeV/$c$ and $3.00 < P_\perp < 3.50$ GeV/$c$. The error bars
  reflect an integrated luminosity of 10~fb$^{-1}$/nucleon.}
\label{fig:combo}
\end{figure} 
Figure \ref{fig:combo} shows the resulting
$\mathrm{d}\sigma/\mathrm{d}\phi$ distributions for the original
parton pairs and the reconstructed dijets  in $\sqrt{s}$=90 GeV $e$+Au collisions for $1.25 <
q_\perp < 1.75$ GeV/$c$ and $3.00 < P_\perp < 3.50$ GeV/$c$. The
results are based on 10M generated events but the error bars were
scaled to reflect an integrated luminosity of 10~fb$^{-1}$/nucleon. The 
plot a) shows the azimuthal anisotropy for all virtual photon
polarizations, and plots b) and c) for transversal and
longitudinal polarized photons, respectively. The quantitative measure
of the anisotropy, $v_2$, is listed in the figures. The values shown
are those for parton pairs; the accompanying numbers in parenthesis
denote the values derived from the reconstructed dijets.
The reconstructed dijets reflect the original anisotropy at the parton
level rather well. The loss in dijet yield is mostly due to  low-$p_T$
particles and is on the order of $25$\%.

While MCDijet provides a tool to study the signal anisotropy in great
detail it does neither generate complete events, nor does it allow us
to derive the level of false identification of dijets in events
unrelated to dijet production. The purity of the extracted signal
sample ultimately determines if these measurements can be conducted.
For studies of this kind we have to turn to PYTHIA6, an event
generator that includes a relatively complete set of DIS processes.

Most of the measurements with dijets   in $e+p$
collisions at HERA (see for example \cite{Aktas:2007bv,
  Gouzevitch:2008zza})  were carried out at
high $Q^2$ and high jet energies ($E_\mathrm{jet} > 10$~GeV). Here, however, we focus on moderately low virtualities and
relatively small jet transverse momenta $P_\perp$. As a result, the dijet signal is easily
contaminated by beam remnants.  To minimize this background source we
limit jet reconstruction to $1 < \eta < 2.5$, sufficiently far away from
the beam fragmentation region.

\begin{figure}[h]
\centering \includegraphics[width=0.5\linewidth]{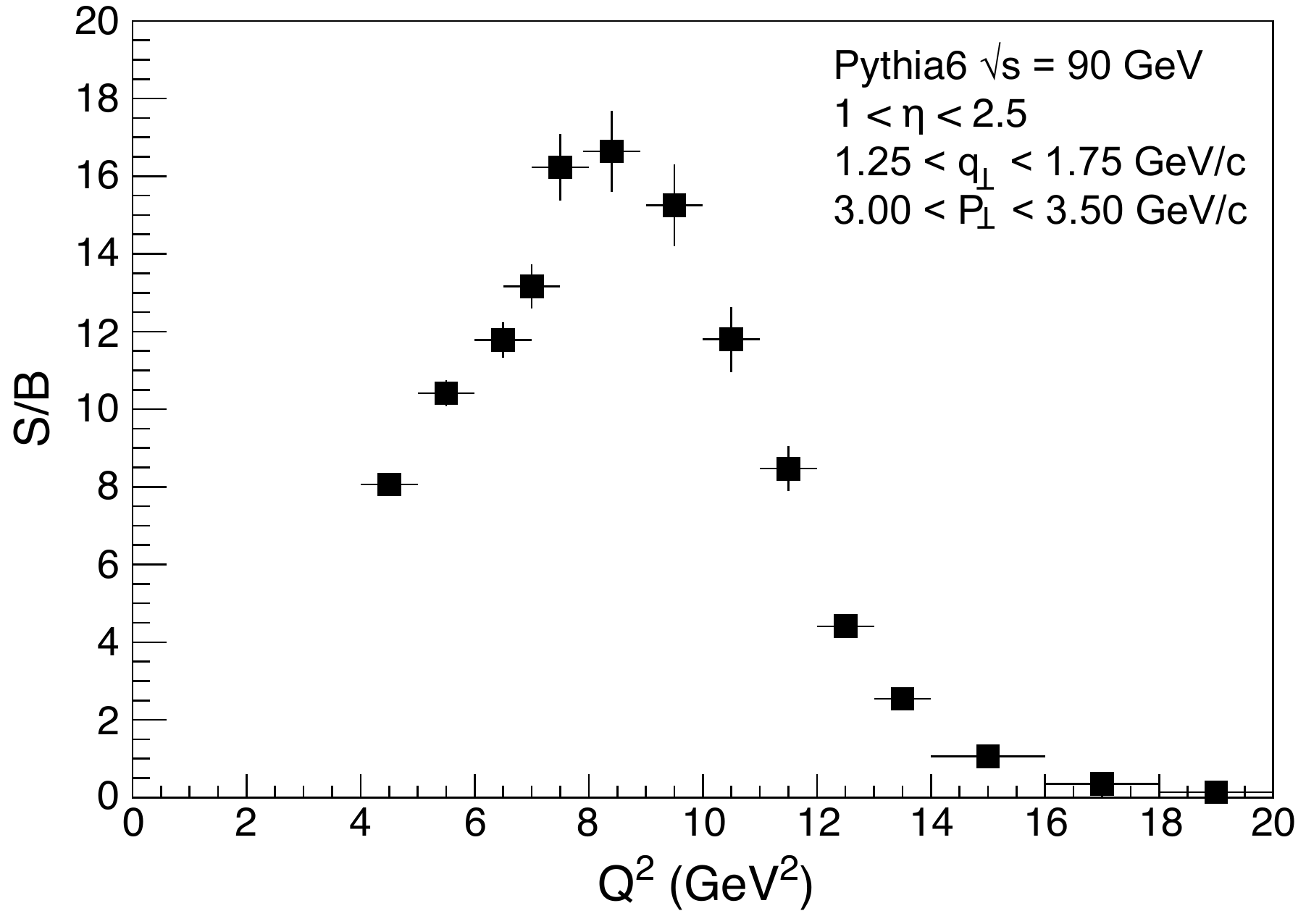}
\caption{$Q^2$ dependence of the signal-to-background ratio derived from PYTHIA6.}
\label{fig:Q2Dep}
\end{figure} 
In PYTHIA6 study, we count $f_i + \gamma^*_\mathrm{T,L} \rightarrow
f_i + g$ and $g + \gamma^*_\mathrm{T,L} \rightarrow f_i + \bar{f_i}$
 as signal and all other as
background processes. The dominant background source is the
standard LO DIS process $ \gamma^* + q \rightarrow q$. Figure
\ref{fig:Q2Dep} illustrates the $Q^2$ dependence of the
signal-to-background ratio, {\it i.e.}, the number of correctly
reconstructed signal events over the number of events that were
incorrectly flagged as containing a signal dijet process. The signal-to-background 
ratio rises initially due to the improved dijet reconstruction
efficiency towards larger $Q^2$ (or $P_\perp$) but then drops
dramatically as particles from the beam remnant increasingly affect
the jet finding. In what follows, we  limit our study to $4
\leq Q^2 \leq 12$ GeV$^2$.

In order to derive the distribution of linearly polarized gluons via
Eqs.~(\ref{eq:v2_L_T}), the contributions from transverse ($v_2^T$)
and longitudinally polarized photons ($v_2^L$) need to be
disentangled. With the exception of diffractive $J/\psi$ production,
no processes in DIS exist where the polarization of the virtual photon
can be measured directly. In our case there are three features that do
make the separation possible:  $v_2^L$ and $v_2^T$ have opposite
signs (see Fig.~\ref{fig:combo}), the background contribution
shows no anisotropy, and the existence of the  relation
\begin{equation}
	v_2^\mathrm{unpol} = \frac{R v_2^L + v_2^T}{1+R}\,, 
	\hspace{2cm} R =
\frac{8 \epsilon_f^2 P_\perp^2\, z (1-z) }{(z^2+(1-z)^2) \, (\epsilon_f^4 +
  P_\perp^4)}\,.
\label{eq:Rformula}
\end{equation}

\begin{figure}[htb]
	\centering \includegraphics[width=0.5\linewidth]{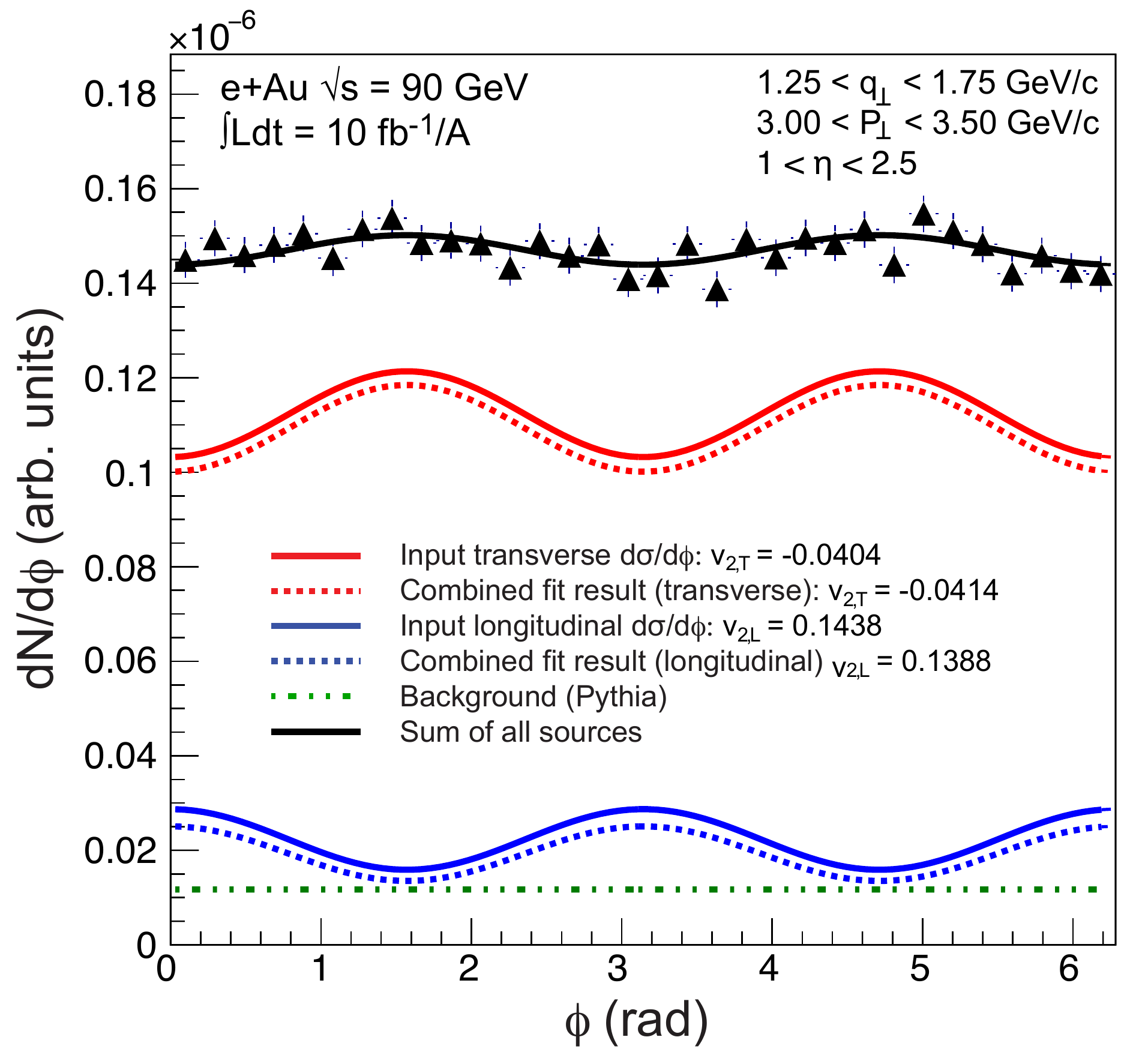}
	\caption{Result of a fit of combined signal and background to a data sample obtained
		in $\sqrt{s}=90$ GeV $e$+A collisions with an integrated luminosity of 10 fb$^{-1}$/nucleon. }
	\label{fig:extract}
\end{figure} 
Our strategy is to perform a combined 5-parameter fit of all 3
components to the full data sample: The signal for longitudinal polarization
($\sigma_L, v_2^L$), that for transverse polarization ($\sigma_T,
v_2^T$), and the flat background ($\sigma_b$). We generated the
data sample in a separate Monte-Carlo combining the signal from
MCDijet with the background contribution from PYTHIA6 while smearing
each data point randomly according to the statistics available at a
given integrated luminosity. The fit provides the desired $v_2^L$ and
$v_2^T$. 

Figure \ref{fig:extract} shows the result of one typical fit on data
generated for a integrated luminosity of 10 fb${^1}$/nucleon. The scatter
and errors on the data points reflect the size of the potential data
sample, the red and the blue curves illustrate the input (solid curve)
and the fit result (dashed curve) for $v_2^L$ and $v_2^T$. The dashed
curves were offset for better visibility.

Additionally systematic studies not presented here showed that the
relative errors improve with increasing $P_\perp$, {\it i.e.},
increasing $v_2$.  Our results indicate that a proper measurement of
the linearly polarized gluon distribution will require integrated
luminosities of at least 20 fb$^{-1}$/nucleon or more. 

\section{Conclusion} 
Monte-Carlo simulations with restrictions of a realistic
detector environment show that it is feasible to study 
the  Weizs\"acker-Williams  transverse momentum dependent (TMD)   gluon  distributions, in particular, linearly polarized distribution, at an electron-ion collider. This, however, might require
a multi-year program assuming that an 
initial EIC luminosity is around $10^{33}$ cm$^{-2}$
s$^{-1}$, as  a proper measurement of
the linearly polarized gluon distribution demands integrated
luminosities of at least 20 fb$^{-1}$/nucleon or more.

\section{Acknowledgement}

I am indebted to  my collaborators, Andrian Dumitru and Thomas Ullrich, for countless hours we spent on stumbling upon, discussing, and, eventually, resolving the problems 
of this project.

I thank E. Aschenauer, J. Huang, and D. Morrison for encouraging me to work on topics related to this project.

\end{document}